\documentclass[aps,showpacs,11pt]{revtex4}
\usepackage{dcolumn}
\usepackage{graphicx}
\usepackage{amsmath}
\usepackage{amsfonts}
\usepackage{amssymb}
\usepackage{psfrag}
\usepackage{wrapfig}
\usepackage{subfigure}
\usepackage{makeidx}
\usepackage{bm}
\usepackage{epsf}
\usepackage{hyperref}
\usepackage{cases}
\makeatletter

\newcommand{\eq}[1]{Eq.(\ref{#1})}

\newcommand{\Rmnum}[1]{\uppercase\expandafter{\romannumeral #1}}

\makeatother

\begin{document}

\title{Critical behaviors of black holes in the Einstein-Maxwell gravity with conformal anomaly}

\author{ Ming Zhang$^{1}$, Rui-Hong Yue$^{2}$\footnote{ Email:rhyue@yzu.edu.cn} and Zhan-Ying Yang$^{1}$}

\affiliation{ $^{1}$School of Physics, Northwest University, Xi'an, 710069, China\\
$^{2}$College of physical science and technology, Yangzhou University, Yangzhou, 225009, China}

\date{\today}

\begin{abstract}
  \indent We study $P-V$ critical behavior of 4-dimensional AdS black hole in the Einstein-Maxwell
  gravity with conformal anomaly by treating the cosmological constant as a variable related
  to the thermodynamic pressure.
  It shows that there is no phase transition if taking $k=0$ or -1.
  When the charge $q_1$ of conformal field and the coefficient $\alpha$ satisfy a certain relation,
  the Van der Waal like phase transition for the spherical black hole can occur
  in case of the temperature is lower than the small
   critical temperature or higher than the large one. We also evaluate the critical exponents of the
  phase transitions and find that the thermodynamic exponents
  associated with this 4-dimensional AdS black hole coincide with
  those of the van der Waals fluid.
\end{abstract}



\maketitle

\section{Introduction}
\label{1s}

Black hole thermodynamics is one of the most interesting objects in physics for many years.
In view of the AdS/CFT correspondence \cite{Maldacena:1997re,Gubser:1998bc,Witten:1998qj},
phase transitions in asymptotically AdS black holes allow for a dual interpretation
in the thermal conformal field theory living on the AdS boundary-the principal example
being the well known radiation/Schwarzschild-Ads black hole Hawking-Page transition \cite{Hawking:1982dh}.

Recently, the analogy between four dimensional RN-AdS black holes and the Van der Waals
fluid-gas system has been completed in this extended phase space \cite{Kubiznak:2012wp},
where the cosmological constant is treated as a thermodynamic pressure
in the geometric units $\hbar=c=1$. There exist some more meaningful reasons to regard the
cosmological constant as a variable \cite{Kubiznak:2012wp}. Firstly, some more fundamental
theories could be considered, where physical quantities, such as Yukawa coupling, gauge coupling constant,
Newton's constant, or cosmological constant may not be fixed values, but can vary arising
from the vacuum expectation energy \cite{Gibbons:1996af,Creighton:1995au}.
Secondly, the first law of black hole thermodynamics cannot be consistent with the Smarr
relation if there is no a variation of $\Lambda$. In the extended phase space, however,
the Smarr relation is satisfied in addition to the first law of thermodynamics from the
aspect of the scaling arguments \cite{Kastor:2009wy}-\cite{YiHuan:2010zz}. Moreover,
the black hole mass $M$ is identified with enthalpy rather than internal energy once
one regards the cosmological constant as thermodynamic pressure in the first law \cite{Kastor:2009wy}.
This analogy becomes more natural recently in the extended phase space. Until now,
these critical behaviour of a lot of black hole systems in this extended phase space
are under discussion in this direction, which are high dimensional charged
black holes \cite{Kubiznak:2012wp,Belhaj:2012bg,Spallucci:2013osa},
rotating black holes \cite{Poshteh:2013pba}-\cite{Altamirano:2013uqa},
Gauss-Bonnet black holes \cite{Wei:2012ui}-\cite{Cai:2013qga}, $f(r)$ black holes \cite{Chen:2013ce}, third-order Lovelock
black holes \cite{Frassino:2014pha,Xu:2014tja} and so on \cite{Mo:2014qsa}-\cite{Liu:2013koa}.
It is worth to noting that, in four-dimensional Born-Infeld AdS black holes \cite{Gunasekaran:2012dq},
the impact of the nonlinearity can bring the new phenomenon of reentrant phase transition
which was observed in rotating AdS holes \cite{Altamirano:2013uqa,Altamirano:2013ane},
while this reentrant phase transition does not occur for higher dimensional Born-Infeld AdS black holes \cite{Zou:2013owa}.

As we known, conformal anomaly,
an important concept with a long history \cite{Duff:1977ay}-\cite{Christensen:1977jc}, has various applications in quantum field theory in curved spaces,
string theory, black hole physics, statistical mechanics and cosmology \cite{Robinson:2005pd}-\cite{Iso:2006wa}.
In quantum field theory in curved spaces, it is required to include the
backreaction of the quantum fields to the spacetime geometry itself
\begin{eqnarray}
R_{\mu\nu}-\Lambda g_{\mu\nu}-\frac{1}{2}Rg_{\mu\nu}=8\pi G<T_{ab}>.\label{1a}
\end{eqnarray}
Cai et al.\cite{Cai:2009ua,Cai:2014jea} have been presented the
(charged) black hole solutions in the gravity
with conformal anomaly in flat and AdS space, whose thermodynamic quantities were also
investigated in the same papers.
In this paper, we would like to focus our attention on the critical behavior of the
black hole with conformal anomaly in AdS space.
As pointed out in Ref \cite{Cai:2009ua}, the solution of the black hole with conformal anomaly reduces to
\begin{eqnarray}
  f(r)=k+\frac{r^2}{l^2}-\frac{2GM}{r}+\frac{\tilde{Q}^2}{r^2},\label{2a}
\end{eqnarray}
where $Q^2=8\pi G (q_1^2+q_2^2)$, $q_1$ is interpreted as $U(1)$ charge of some conformal
field theory and $q_2$ is the charge of Maxwell field.
Note that this solution (\eq{2a}) is similar with the form of the
Reissner-Nordstr$\ddot{o}$m-AdS black hole solution.
The integration constant $q_1$ in the solution is nothing but a ``dark radiation'' term
which cannot be excluded. However, under this condition, $q_1$ is undistinguished
from the electric charge of the electric field. So only the charge ($q_2$) of the Maxwell field left.
Until now, $P-V$ critical behaviors of the RN-AdS black hole has been analyzed in \cite{Kubiznak:2012wp}.
What about the role that the conformal anomaly term plays in phase transition ?
It is interesting to discuss.
We will find that no phase transition happens for $k=0$ or -1;
when the charge $q_1$ of conformal field and the coefficient $\alpha$
satisfy a certain relation, the Van der Waal like phase transition for the spherical black hole can occur
when the temperature is lower than the small critical temperature or higher than the large one.

This paper is organized as follows. In Sec.~\ref{2s}, we present the phase transition of the
conformal anomaly corrected black holes. The Sec.~\ref{4s} is devoted to the closing remarks.

\section{Conformal anomaly corrected black hole solution in AdS spacetimes}
\label{2s}

The static, spherically symmetric black hole solution in Einstein-Maxwell gravity
theory with conformal anomaly has been presented by  \cite{Cai:2013qga}
\begin{eqnarray}
&&ds^2=-f(r)dt^2+\frac{1}{f(r)}dr^2+r^2d\Omega_{2k}^2,\label{eq:1a}\\
&&f(r)=k-\frac{r^2}{4\tilde{\alpha}}\left(1-\sqrt{1+\frac{8\tilde{\alpha}}{l^2}
-\frac{16\tilde{\alpha}G M}{r^3}
+\frac{64\pi G\tilde{\alpha}(q_1^2+q_2^2)}{r^{4}}}\right),\label{eq:2a}
\end{eqnarray}
where the integration constants $M$ is nothing but the mass of the solution
while $q_1$ should be interpreted as $U(1)$ charge of some conformal
field theory and $q_2$ is the charge of Maxwell field.
The parameter $\tilde{\alpha}$ is defined as $\tilde{\alpha}=8\pi G \alpha_1$ and $\alpha_1$
is a positive constant related to the content of the conformal field theory.
Moreover, $d\Omega_{2k}^2$ is the line element of a two-dimensional Einstein constant
curvature space with scalar curvature $2k$. Here
the constant $k$ characterizes the geometric property of
hypersurface and takes values $k=0$ (flat), $k=-1$
(negative curvature) and $k=1$ (positive curvature).

In terms of horizon radius $r_+$, the mass $M$ of black hole reads as
\begin{eqnarray}
M=\frac{Q^2}{2Gr_+}+\frac{kr_+}{2G}-\frac{k^2\tilde{\alpha}}{Gr_+}+\frac{r_+^3}{2Gl^2}.\label{eq:3a}
\end{eqnarray}
The Hawking temperature $T$ can be derived as
\begin{eqnarray}
T=\frac{1}{4\pi}\frac{df(r_+)}{dr_+}=\frac{3r_+^4+l^2\left(k(r_+^2+2k\tilde{\alpha})-Q^2\right)}{4\pi l^2\left(r_+^3
-4k\tilde{\alpha}r_+\right)}.\label{eq:4a}
\end{eqnarray}
We try to obtain the black hole entropy $S$ by employing
the first law of black hole thermodynamics with
\begin{eqnarray}
S=\int T^{-1}\left(\frac{\partial M}{\partial r_+}\right)dr_+=\frac{\pi r_+^2}{G}
-\frac{4k\pi\tilde{\alpha}}{G}\ln(r_+^2)+S_0.\label{eq:5a}
\end{eqnarray}
Here $S_0$ is an integration constant, which unfortunately we cannot fix because of the existence
of the logarithmic term.

Here we discuss the critical behavior of black holes in the extended phase
space, where the cosmological constant $\Lambda$ is identified with thermodynamic pressure $P$,
\begin{eqnarray}
P=-\frac{\Lambda}{8\pi}=\frac{3}{8\pi l^2}\label{eq:6a}
\end{eqnarray}
in the geometric units $G_N=\hbar=c=k=1$.
Then we find that those thermodynamic quantities satisfy the following differential form
\begin{eqnarray}
dM=TdS+\Phi_1 dq_1+\Phi_2 dq_2+VdP+\Omega d\tilde{\alpha},\label{eq:8a}
\end{eqnarray}
where $V$ and $\Phi$ denote the thermodynamic volume and electric potential (chemical potential) with
\begin{eqnarray}
&&V=\left(\frac{\partial M}{\partial P}\right)_{S,\Phi,\tilde{\alpha}}=\frac{4\pi r_+^{3}}{3},
\quad \Omega=\left(\frac{\partial M}{\partial\tilde{\alpha}}\right)_{S,\Phi,P}
=-\frac{k^2}{Gr_+}.\label{eq:9a} \\
&&\Phi_1=\frac{8\pi}{r_+}q_1, \quad \Phi_2=\frac{8\pi}{r_+}q_2.
\end{eqnarray}

By the scaling argument, we can obtain the generalized Smarr relation
\begin{eqnarray}
  M=2TS-2PV+2\Omega\tilde{\alpha}+\Phi_1q_1+\Phi_2q_2.
\end{eqnarray}

\section{Critical behaviors of conformal anomaly corrected black holes in AdS spacetimes}
\label{3s}

With Eqs.~(\ref{eq:4a})(\ref{eq:6a}), the equation of state $P(V,T)$ is obtained
\begin{eqnarray}
P=\frac{T}{2 r_+}\left(1-\frac{4k\tilde{\alpha}}{r_+^2}\right)-\frac{k}{8\pi r_+^2}
-\frac{k^2\tilde{\alpha}}{4\pi r_+^4}+\frac{Q^2}{8\pi r_+^4}.\label{eq:10a}
\end{eqnarray}
To make contact with the Van der Waals fluid equation in $4$-dimensions, translating the ``geometric" equation
of state to physical one by identifying the specific volume $v$ of the fluid with
the horizon radius of the black hole as $v=2r_+$.

We know that the critical points are determined as the inflection in the $P-r_+$ diagram, as
\begin{eqnarray}
\frac{\partial P}{\partial r_+}\Big|_{T=T_c, r_+=r_c}
=\frac{\partial^2 P}{\partial r_+^2}\Big|_{T=T_c, r_+=r_c}=0.\label{eq:11a}
\end{eqnarray}
Then we can obtain the critical temperature
\begin{eqnarray}
T_c=\frac{4k^2\tilde{\alpha}+k r_c^2-2Q^2}{2\pi r_c\left(r_c^2-12k\tilde{\alpha}\right)} \label{eq:12a}
\end{eqnarray}
and the equation for the critical horizon radius $r_c$ (specific volume $v_c=2r_c$) is
\begin{eqnarray}
  k r_c^4+6\big(4\tilde{\alpha}k^2-Q^2\big)r_c^2-24\tilde{\alpha}k(2\tilde{\alpha}k^2-Q^2)=0, \label{eq:13a}
\end{eqnarray}
where $r_c$ denotes the critical value of $r_+$.
Obviously the critical temperature $T_c$ (Eq.~(\ref{eq:12a}))
is negative for $k=0$, and then the Van de Waals like SBH/LBH phase transition does not happens.
So, we only explore the cases of $k=-1$ and $k=1$.

\subsubsection{$k=-1$}

For $k=-1$, the root of Eq.~(\ref{eq:13a}) is given by
\begin{eqnarray}
r_{c}^2=3\left(4\tilde{\alpha}-Q^2\right)\pm\sqrt{3(Q^2-8\tilde{\alpha})(3Q^2-8\tilde{\alpha})} \label{eq:14a}
\end{eqnarray}
in case of $Q^2\geq8\tilde{\alpha}$ or $Q^2\leq\frac{8\tilde{\alpha}}{3}$. Furthermore,
the positivity of $r_{c}^2$ makes the condition $Q^2\leq\frac{8}{3}\tilde{\alpha}$
to be satisfied for the ``+" branch
and the condition $2\tilde{\alpha}<Q^2\leq\frac{8\tilde{\alpha}}{3}$ for the ``$-$" branch.
Considering the different branch of $r_c^2$ when $Q$ and $\tilde{\alpha}$ are located
in corresponding region, however, we find that the critical temperature $T_c$ (Eq.(\ref{eq:12a}))
always maintains negative, and then the Van der Waals like SBH/LBH
phase transition does not happen for the hyperbolic black hole.

\subsubsection{$k=+1$}

From Eq.~(\ref{eq:13a}), we can obtain
\begin{eqnarray}
r_{c}^2=3\left(Q^2-4\tilde{\alpha}\right)\pm\sqrt{3(Q^2-8\tilde{\alpha})(3Q^2-8\tilde{\alpha})}\label{eq:15a}
\end{eqnarray}
Here we only take ``$-$" branch when $Q^2\geq8\tilde{\alpha}$
and positive one with ``$+$" branch for $Q^2\geq8\tilde{\alpha}$
or $Q^2<2\tilde{\alpha}$ because of $r_c>0$. In other words, Eq.~(\ref{eq:13a}) could admit two
solutions of $r_c$ if taking $Q^2\geq8\tilde{\alpha}$, while only admits one positive root
when $Q^2<2\tilde{\alpha}$.

\begin{figure}[htb]
  \subfigure[$P-r_+$]{\label{fig:subfig:a} 
  \includegraphics{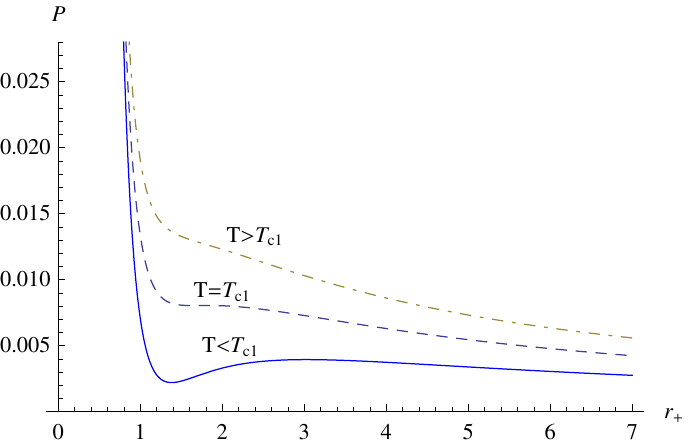}}%
  \hfill%
  \subfigure[$F-T$]{\label{fig:subfig:b} 
  \includegraphics{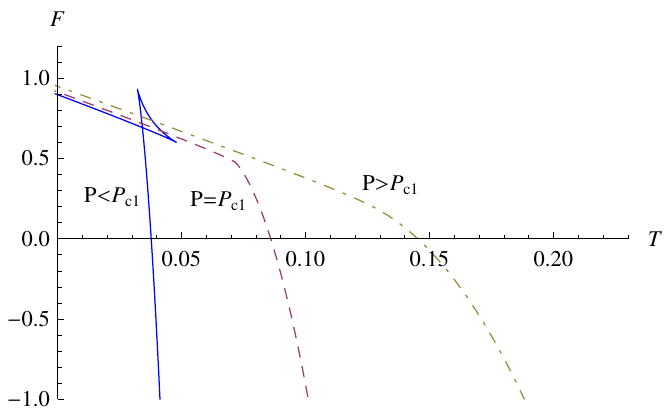}}%
  \caption{ The $P-r_+$ and $F-T$ of conformal anomaly corrected black holes
  with $Q=1$ and $\tilde{\alpha}=0.1$ } near the critical point 1.
\end{figure}

\begin{figure}[htb]
  \subfigure[$P-r_+$]{\label{fig:subfig1:a} 
  \includegraphics{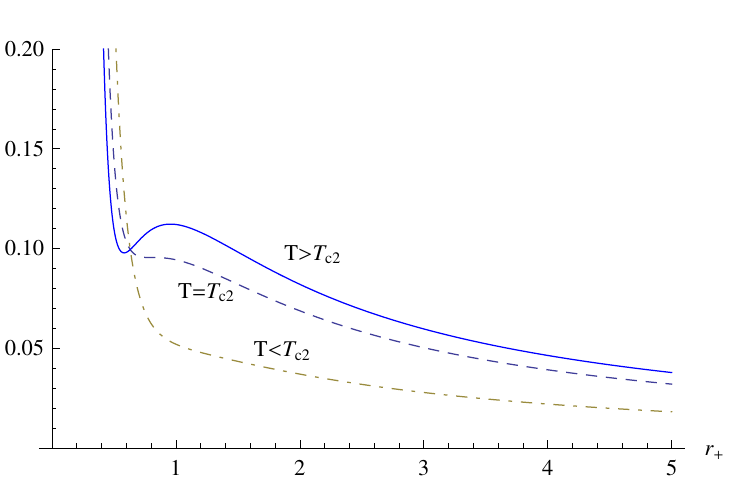}}%
  \hfill%
  \subfigure[$F-T$]{\label{fig:subfig1:b} 
  \includegraphics{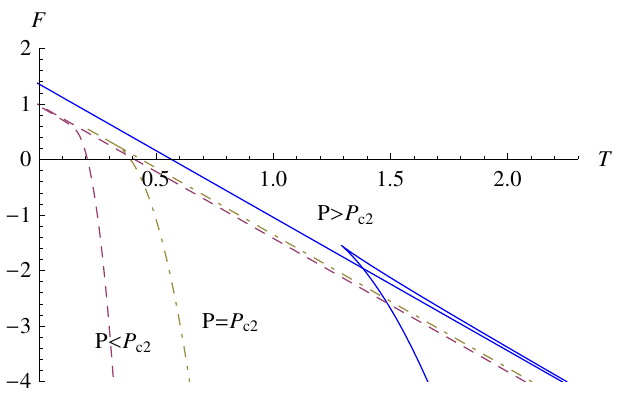}}%
  \caption{ The $P-r_+$ and $F-T$ of conformal anomaly corrected black holes
  with $Q=1$ and $\tilde{\alpha}=0.1$ } near the critical point 2.
\end{figure}

Then, we can also continue to discuss the critical temperature $T_c$ (Eq.~(\ref{eq:12a})) and
pressure $P_c$ (Eq.~(\ref{eq:10a})), and find that
the values of $T_c$ and $P_c$ are positive in case of $Q^2\geq8\tilde{\alpha}$,
which implies there exist two critical points. Nevertheless,
$T_c$ and $P_c$ always take negative values when $Q^2<2\tilde{\alpha}$.
We plot the $P-r_+$ isotherm diagram of the spherical black hole in Fig.\ref{fig:subfig:a}
and Fig.\ref{fig:subfig1:a}, which depict respectively the critical behavior near the
critical point 1 (the ``$+$'' branch) and the critical point 2 (the ``$-$'' branch).
We can find that these two diagrams are exactly the same as the $P-V$ diagram of the Van der Waals liquid-gas system.
For $Q^2>8\tilde{\alpha}$, Fig.1a shows that the two upper dashed lines correspond to
the ``idea gas" phase behavior when $T>T_{c1}$, and the Van de Waals like SBH/LBH phase
transition appears in the system when $T<T_{c1}$. While in Fig.~2a, For a fixed temperature higher than the critical
one $T_{c2}$, we have two branches whose pressure decreases as the increase of horizon radius,
one is in the small radius region (corresponding to fluid phase) and the other is in the large
radius region (corresponding to the gas phase). However, the black holes are always in the
gas phase and no phase transition happens behind the critical temperature $T_{c2}$.
and the Van de Waals like SBH/LBH phase transition appears in the system when $T>T_{c2}$.
We can also verify that the first critical temperature $T_{c1}$ is
lower that the second critical temperature $T_{c2}$ in the region $Q^2>8\tilde{\alpha}$.

In addition, the behavior of free energy $F$ is important to investigate the thermodynamic phase
transition of conformal anomaly corrected black holes. In fixed a charge $Q$, the
free energy $F$ reads as
\begin{eqnarray}
  F=M-TS\label{eq:22a}
\end{eqnarray}
Here $r_+$ is understood as the function of pressure and temperature, $r_+=r_+(P,T)$,
via equation of state Eq.~(\ref{eq:15a}). As show in Fig.~1b and Fig.~2b, these figures develops
a ``swallow tail" for $P<P_{c1}$ and $P>P_{c2}$, which denote a first order transition and this ``swallow tail" vanishes at $P_{c1}\le P$ and $ P\le P_{c2}$.

\section{Critical exponents near critical point}
\label{4s}

Now we turn to compute the critical exponents $\alpha$, $\beta$, $\gamma$, $\delta$ characterizing the behavior
of physical quantities in the vicinity of the critical point $(r_+=r_c, v=v_c, T=T_c, P=P_c)$ for
the black hole in the gravity with conformal anomaly in four-dimension.
Near the critical point, the critical behavior of a Van de Waals liquid-gas system can be
characterized by the following critical exponents \cite{Kubiznak:2012wp}.
\begin{eqnarray}
&&C_v=T\frac{\partial S}{\partial T}\Big|_v\propto \left(-\frac{T-T_c}{T_c}\right)^{-\alpha},\nonumber\\
&&\eta=\frac{v_s-v_l}{v_c}\propto \left(-\frac{T-T_c}{T_c}\right)^{\beta},\nonumber\\
&&\kappa_T=-\frac{1}{v}\frac{\partial v}{\partial P}\Big|_T\propto \left(-\frac{T
-T_c}{T_c}\right)^{-\gamma},\nonumber\\
&& P-P_c \propto (v-v_c)^{\delta},\label{eq:27a}
\end{eqnarray}
where $``c"$ denotes that a quantity is taken values at the critical point of
the Van de Waals liquid-gas system.

In order to compute the critical exponent $\alpha$, the entropy $S$ of black
hole (Eq.~(\ref{eq:5a})) can be rewritten as
$\frac{\pi r_+^2}{G}-\frac{4\pi\tilde{\alpha}}{G}\ln(r_+^2)+S_0$.
Obviously this entropy $S$ is independent of $T$ with the constant values of specific volume $v$,
and then we get the critical exponent $\alpha=0$. To obtain other exponents, defining
\begin{eqnarray}
p=\frac{P}{P_c}, \quad \nu=\frac{v}{v_c}, \quad \tau=\frac{T}{T_c}, \label{eq:28a}
\end{eqnarray}
we introduce the expansion parameters
\begin{eqnarray}
\tau=t+1, \quad \nu=\omega+1,\label{eq:29a}
\end{eqnarray}
then the expansion for this equation of state near the critical point is given by
\begin{eqnarray}
p=1+a_{10}t+a_{11}t\omega+a_{03}\omega^3+\mathcal{O}(t\epsilon^2,\epsilon^4).\label{eq:30a}
\end{eqnarray}
We obtain all the expansion coefficients in Eq.~(\ref{eq:30a}) through computing its derivatives
with respect to $t$ and $\omega$ at the critical point. These coefficients are obtained
\begin{subequations}
  \begin{numcases}{}
    a_{10}\approx 2.21, \quad a_{11}\approx-1.51, \quad a_{03}\approx-1.45. ~(\text{critical point 1})\label{eq:31a} \\
    a_{10}\approx 0.853, \quad a_{11}\approx 1.87, \quad a_{03}\approx-1.79. ~~(\text{critical point 2})\label{eq:31b}
  \end{numcases}
\end{subequations}
For $T<T_{c1}$ or $T>T_{c2}$, there exist two different volumes for a same pressure during the phase transition.
Then we have
\begin{eqnarray}
&&p=1+a_{10}t+a_{11}t\omega_s+a_{03}\omega_s^3=1+a_{10}t+a_{11}t\omega_l+a_{03}\omega_l^3, \nonumber\\
\Rightarrow && a_{11}t\left(\omega_s-\omega_l\right)+a_{03}\left(\omega_s^3-\omega_l^3\right)=0,\label{eq:32a}
\end{eqnarray}
where $\omega_s$ and $\omega_l$ denote the `volume' of small and large black holes.

In addition, using Maxwell's area law, we can obtain
\begin{eqnarray}
\int^{\omega_s}_{\omega_l}\omega\frac{dp}{d\omega}d\omega=0 \Rightarrow a_{11}t\left(\omega_s^2-\omega_l^2\right)
+\frac{3}{2}a_{03}\left(\omega_s^4-\omega_l^4\right)=0.\label{eq:33a}
\end{eqnarray}
With Eqs.~(\ref{eq:32a})(\ref{eq:33a}), the nontrivial solutions appear only when $a_{11}a_{03}t<0$.
Then we can obtain
\begin{eqnarray}
\omega_s=\frac{\sqrt{-a_{11}a_{03}t}}{3|a_{03}|}, \quad
\omega_l=-\frac{\sqrt{-a_{11}a_{03}t}}{3|a_{03}|}.\label{eq:34a}
\end{eqnarray}
Therefore, we have
\begin{eqnarray}
\eta=\omega_s-\omega_l=2\omega_s=\frac{2\sqrt{-a_{11}a_{03}}}{3|a_{03}|}\sqrt{t}\Rightarrow \beta=1/2.\label{eq:35a}
\end{eqnarray}

The isothermal compressibility can be computed as
\begin{eqnarray}
\kappa_T=-\frac{1}{V}\frac{\partial V}{\partial P}\Big|_{V_c}\propto
-\frac{1}{\frac{\partial p}{\partial \omega}}\Big|_{\omega=0}=\frac{2}{3t},\label{eq:36a}
\end{eqnarray}
which indicates that the critical exponent $\gamma=1$. Moreover, the shape of the critical
isotherm $t=0$ is given by
\begin{eqnarray}
p-1=-\omega^3\Rightarrow \delta=3.\label{eq:37a}
\end{eqnarray}
Evidently we can see that for critical point 1 or critical point 2, these critical exponents of the
black hole in the gravity with conformal anomaly coincide with those of the Van der Waals liquid-gas system.

\section{Closing remarks}
\label{5s}

In this paper we have studied the phase transition and critical behavior of 4-dimensional
black hole in the gravity with conformal anomaly, where the cosmological constant is treated
as a variable related to the thermodynamic pressure.
For $k=0,-1$, there is no critical behavior in the system.
It is interesting that for $k=+1$ we find under the constraint condition $Q^2>8\tilde{\alpha}$,
there exist two critical point in the system and two phase transitions when the temperature of the
black hole monotonically varied. The crucial feature in our case is that the two phase transitions
occur when the temperature is lower that the small critical temperature ($T<T_{c1}$) and higher that
the large one ($T>T_{c2}$). Furthermore, We calculate the corresponding critical exponents of the two
SBH/LBH phase transitions, and the results are in accordance with Van der Waals fluid system.

In Ref.\cite{Kubiznak:2012wp}, the author found that there exists only one critical point,
for 4-dimensional RN-AdS black hole with $k=+1$, and then
the system demonstrates the Van der Waals like small/large black hole
phase transition occur when the temperature is
lower that the critical temperature. Obviously, the conformal anomaly term brings
richer phase structures and critical behaviors than that of RN-AdS black hole.

{\bf Acknowledgements}

We would like to thank Dr. Zou De-Cheng and Dr. Xu Wei for many discussions and a careful reading of this paper.
This work was supported by the National Natural Science Foundation of China under Grant Nos.11275099 and 11475135.

\end{document}